\documentstyle[12pt,epsfig]{article}

\def\be{\begin{equation}}
\def\ee{\end{equation}}

\begin{document}

\begin{titlepage}
\setlength{\textwidth}{5.0in}
\setlength{\textheight}{7.5in}
\setlength{\parskip}{0.0in}
\setlength{\baselineskip}{18.2pt}
\setlength{\footskip}{0.5in}
\setlength{\footheight}{0in}

\renewcommand{\thefootnote}{\fnsymbol{footnote}}

\hfill SNUTP 99-057\\
\vspace{0.3cm}
\hfill SOGANG-HEP 264/99

\vspace{0.3cm}

\begin{center}
{\Large\bf Reissner--Nordstr\"om--AdS black hole\\
in the GEMS approach}
\end{center}

\begin{center}
{Yong-Wan 
Kim\footnote{Electronic address: ywkim65@netian.com}$^{1,2}$,
Young-Jai 
Park$\footnote{Electronic address: yjpark@ccs.sogang.ac.kr}^{3}$,
and Kwang-Sup 
Soh\footnote{Electronic address: kssoh@phya.snu.ac.kr}$^{2}$}\par
\end{center}

\begin{center}
{${}^{1}$JeungSanDo Research Institute, Sonhwadong 356-8,}\par 
{Jung-gu Taejeon, 301-050, Korea,}\par
{${}^{2}$Department of Physics Education
and Center for Theoretical Physics,}\par
{Seoul National University, Seoul 151-742, Korea,}\par
{and }\par
{${}^{3}$Department of Physics and Basic Science Research Institute,}\par
{Sogang University, C.P.O. Box 1142, Seoul 100-611, Korea}\par

\vskip 0.5cm
{\today}
\end{center}
\vfill

\begin{center}
{\bf ABSTRACT}
\end{center}
We obtain a (5+2)-dimensional global flat embedding of the 
(3+1)-dimensional curved RN--AdS space. 
Our results include the various limiting cases of
global embedding Minkowski space (GEMS) geometries
of the RN, Schwarzschild--AdS in (5+2)-dimensions, 
Schwarzschild in (5+1)-dimensions, purely charged space, 
and universal covering space of AdS in (4+1)-dimensions,
through the successive truncation procedure of parameters
in the original curved space.

\vskip 0.5cm
\noindent
PACS number(s): 04.70.Dy, 04.20.Jb, 04.62.+v
\end{titlepage}

\newpage

\section{Introduction}

Ever since the discovery that thermodynamic properties of black holes 
in anti--de Sitter (AdS) space-time are dual to those of a field
theory in one dimension fewer, there has been of much interest in
Reissner--Nordstr\"om (RN)--AdS black hole \cite{rnads}, 
which now becomes a prototype example to study this AdS/CFT 
correspondence \cite{wit}. 
On the other hand, after Unruh's work \cite{unr}, it has been known 
that a thermal Hawking effect on a curved manifold \cite{hawking} 
can be looked at as an Unruh effect in a higher dimensional flat 
space-time. Recently, non-trivial works of isometric embeddings 
of the RN black hole \cite{des-prd} and M2-, D3-, M5-branes 
\cite{gibbons} into flat spaces  with two times has been
studied to get some insight of the global aspect of the space-time 
geometries in the context of brane physics.
Moreover, several authors \cite{des, bec, gon} have also 
shown that global embedding Minkowski space (GEMS) approach 
\cite{kasner,fro,goenner,ros,n} of which a hyperboloid in a higher
dimensional space corresponds to original curved space
could provide a unified derivation of temperature for a wide 
variety of curved spaces. These include the static, rotating, charged BTZ 
\cite{btz,kps,hongkp}, the Schwarzschild \cite{sch} together with 
its AdS extensions, and the RN \cite{rn} black holes.
Therefore, it is interesting to study the geometry of the RN--AdS 
and their thermodynamics \cite{pel} in this GEMS approach.

In this paper we will analyze the Hawking and Unruh effects of the
$D=4$ RN--AdS space, which has not been tackled up to now due to
the complicated structure of this system, in terms of the GEMS
approach covering the usual Kruskal extension \cite{kru}. 
In Sec.2, we discuss the $D=4$ RN--AdS embedding into a seven
dimensional flat space. 
In Sec.3, we show that our results in the GEMS
of the RN--AdS space systematically include those of the various
limiting GEMS geometries, which are the RN, Schwarzschild--AdS,
Schwarzschild, purely charged and AdS space-times, 
through the successive
truncation procedure of parameters in the original curved space.
These correspond to the dimensional reduction in the GEMS approach.
Finally, we present summary in Sec.4.

\section{Geometric Structure of RN-AdS \\ in the GEMS Approach}

Let us consider the line element of the four dimensional RN--AdS
space$\footnote{We restrict our discussion to the non-extremal
case.}$ \cite{rnads}
\be
\label{metric}
ds_4^2= f(r,m,e,R)dt^2 -f^{-1}(r,m,e,R) dr^2
             -r^2(d\theta^2+\sin^2\theta d\phi^2),
\ee
where $f(r,m,e,R)$ is given by
\be
f(r,m,e,R) = 1-\frac{2m}{r}+\frac{e^{2}}{r^{2}} + \frac{r^{2}}{R^{2}}.
\ee
This space-time is asymptotically described by AdS,
and there is an outer horizon at $r=r_H$. The case of $e=0$ yields the
Schwarzschild--AdS metric, the case of $m=e=0$ yields the metric on
the universal covering space of AdS \cite{HE},
the case of $R\rightarrow \infty$ yields the RN metric,
and the case of $m=0$ and $R\rightarrow \infty$ yields the
purely charged metric.

To embed this space-time into a higher dimensional flat one, 
we first note that by introducing three coordinates 
$(z^3, z^4, z^5)$ in Eq. (\ref{rna}) (see below) the last term 
in the metric (\ref{metric}) can be written to give 
$-(dz^3)^2-(dz^4)^2-(dz^5)^2=-dr^2-r^2(d\theta^2+\sin^2\theta d\phi^2)$.
Then, making use of an ansatz of two coordinates ($z^0$, $z^1$) 
in Eq. (\ref{rna}), we have obtained 
\begin{eqnarray}
\label{temp}
&& (dz^0)^2-(dz^1)^2-(dz^3)^2-(dz^4)^2-(dz^5)^2 \nonumber \\
&& = f(r,m,e,R)dt^2 - r^2(d\theta^2+\sin^2\theta d\phi^2)
   - \left( 1 + \frac{(\frac{f'(r,m,e,R)}{2})^2}{k^2_H f(r,m,e,R)}\right) 
     dr^2, \nonumber \\
\end{eqnarray}
where $f'(r,m,e,R)$ denotes the derivative with respect to $r$ 
and 
\be
\label{sgrna}
k_H(r_H,e,R^2)=\frac{(r_H^2-e^2 +\frac{3r_H^4}{R^2})}{2r_H^3} > 0
\ee
is the surface gravity at the root of $f(r,m,e,R)\mid_{r=r_H}=0$.
In order to make the form of $ds^2_4$ in Eq. (\ref{metric}),
we subtract the $f^{-1}(r,m,e,R)dr^2$ term from Eq. (\ref{temp}) 
on the right-hand side and add it again to Eq. (\ref{temp}).
Then, the remaining extra radial part of
\be
f^{-1}(r,m,e,R)dr^2 
  - \left( 1 + \frac{(\frac{f'(r,m,e,R)}{2})^2}{k^2_H f(r,m,e,R)}\right) dr^2
\ee
can be seperated into positive and negative definite parts
with $r > r_H$ as follows:
\begin{eqnarray}
&& R\left( \frac{e^2}{[rr_H(r^2+rr_H+r_H^2)+(rr_H-e^2)R^2]} 
    \right.\nonumber\\
&& + \left.\frac{r^2_H (r^2 + rr_H + r_H^2)
    [(r_H^2-e^2)^2 R^4 + r_H^6 (r_H^2+2 R^2)]}
    {r^2 [3r_H^4 + (r_H^2-e^2)R^2]^2
    [rr_H(r^2 + rr_H + r_H^2)+(rr_H-e^2)R^2]}
              \right) dr^2 \nonumber\\
&& - \left(e^2 \frac{R^4 r_H^6 [4(r r_H-e^2)R^2+10r^4+2rr_H(r^2+rr_H+2r^2_H)]}
   {r^4[3r_H^4 + (r_H^2-e^2)R^2]^2[rr_H(r^2+rr_H+r_H^2)+(rr_H-e^2)R^2]}
   \right)dr^2 \nonumber \\
&& -\left( \frac{r r_H(r^2+rr_H+r_H^2)(4r_H^6 R^2+[3r_H^4+(r_H^2-e^2)R^2]^2)}
       {[3r_H^4+(r_H^2-e^2)R^2]^2[rr_H(r^2+rr_H+r_H^2)+(rr_H-e^2)R^2]}
   \right)dr^2 \nonumber \\
&& = (dz^2)^2 - (dz_e)^2 - (dz_R)^2,
\end{eqnarray}
where we have used the relation between the Arnowitt--Deser--Misner
mass of the RN--AdS black hole and its event horizon radius $r=r_H$,
{\it i.e.}, $2m=r_H+r^3_H/R^2+e^2/r_H$.
At this stage, it should be note that due to the existence of the last two terms, 
$e$-sensitive $(dz_e)^2$ and $R$-dominant $(dz_R)^2$, one may think that superficially two 
additional time dimensions are needed for a global flat embedding. 
However, it is in fact enough to introduce only one time 
dimension $(dz^6)^2$ by combining these two terms as
\begin{equation}
(dz^6)^2 = (dz_e)^2 + (dz_R)^2,
\end{equation}
for a desired minimal GEMS\footnote{In the region of $r > r_H$, 
it can be easily verified that the $(dz^2)^2$ and $(dz^6)^2$ 
are positive definite functions, when combined with the condition in Eq. (\ref{sgrna}).}   
 with an additional spacelike dimension $(dz^2)^2$.
Note also that the $(dz_e)^2$ (or, $(dz_R)^2$) term is shown to be vanished 
in the limit of $e\rightarrow 0$ (or, $R\rightarrow \infty$), 
and the $(dz^6)^2$ becomes $(dz_R)^2$ (or, $(dz_e)^2$).
As a result, we have obtained a flat global embedding in (5+2)-dimensions of the 
corresponding curved 4-metric as
\begin{eqnarray}
ds_7^2&=&(dz^0)^2- \sum^5_{i=1} (dz^i)^2 +(dz^6)^2  \label{eight} \\
&=& f(r,m,e,R)dt^2 -f^{-1}(r,m,e,R) dr^2-r^2(d\theta^2+\sin^2\theta d\phi^2) \nonumber\\
&=& ds_4^2.
\end{eqnarray}
This equivalence between the (5+2)-dimensional flat embedding space and 
original curved space is the very definition of isometric embedding, mathematically developed
by several authors \cite{gibbons,nash}. 

It seems appropriate to comment on the lowest embedding dimensions
in terms of the number of parameters.
It is known from the previous works \cite{des-prd,ros,hongkp} 
that whenever one parameter is increased in the original space, 
the embedding dimensions are either unchanged or increased depending on
this parameter. In particular, the embedding dimension is already $D=7$ 
for the case of the RN or Schwarzschild--AdS, which have one less parameters 
than those of the RN--AdS case. Therefore, for the case of the RN--AdS 
the possibly lowest embedding dimension is $D=7$.

In summary, through the GEMS approach which makes the curved space 
possibly embedded in a higher dimensional flat space 
\cite{kasner,fro,goenner,ros,n},
we have found a $D=7$ dimensional isometric embedding of the RN--AdS space
as
\begin{eqnarray}
\label{rna}
&& z^0=k_H^{-1}\sqrt{f(r,m,e,R)}\sinh (k_Ht),\nonumber\\
&& z^1=k_H^{-1}\sqrt{f(r,m,e,R)}\cosh (k_Ht),\nonumber\\
&& z^2=\int dr R
        \left(
              \frac{e^2}{[rr_H(r^2+rr_H+r_H^2)+(rr_H-e^2)R^2]}
         \right. \nonumber \\
&& ~~~  + \left.
\frac{r^2_H (r^2 + rr_H + r_H^2)
            [(r_H^2-e^2)^2 R^4 + r_H^6 (r_H^2+2 R^2)]}
        {r^2 [3r_H^4 + (r_H^2-e^2)R^2]^2
           [rr_H(r^2 + rr_H + r_H^2)+(rr_H-e^2)R^2]}
              \right)^{1/2},   \nonumber \\
&& z^3=r\sin\theta\cos\phi, \nonumber\\
&& z^4=r\sin\theta\sin\phi, \\
&& z^5=r\cos\theta, \nonumber \\
&& z^6=\int dr \left(
  \frac{e^2 R^4 r_H^6 [4(r r_H-e^2)R^2+10r^4
        +2rr_H(r^2+rr_H+2r^2_H)]}
       {r^4[3r_H^4 + (r_H^2-e^2)R^2]^2
       [rr_H(r^2+rr_H+r_H^2)+(rr_H-e^2)R^2]}
       \right.  \nonumber \\
&& ~~~+ \left.
       \frac{r r_H(r^2+rr_H+r_H^2)
        (4r_H^6 R^2+[3r_H^4+(r_H^2-e^2)R^2]^2)}
       {[3r_H^4+(r_H^2-e^2)R^2]^2
        [rr_H(r^2+rr_H+r_H^2)+(rr_H-e^2)R^2]}
       \right)^{1/2}, \nonumber
\end{eqnarray}
with an additional spacelike $z^2$ and a timelike $z^6$ dimensions. 
Therefore, the (3+1)-dimensional curved space is seen as 
the hyperboloid embedded in a (5+2)-dimensional flat space.
It would be easily verified inversely that the flat metric (\ref{eight}) in the  
(5+2)-dimensional space defined as the coordinates (\ref{rna}) 
gives the original RN--AdS metric (\ref{metric}) correctly.

Now, following the trajectory of $z^2=\cdot\cdot\cdot =z^6=0$ in Eq.
(\ref{rna}) which corresponds to a static trajectory
($r,\theta,\phi={\rm constant}$) in the curved space,
the relevant $D=7$ acceleration $a_7$ is described as the
Rindler-like motion \cite{des,n,hongkp} of the form of 
$(z^1)^2-(z^0)^2=a^{-2}_7$ in the embedded flat space, {\it i.e.},
\be
\label{acc8}
a_{7}=\{(z^1)^2-(z^0)^2\}^{-1/2}=
\frac{r_H^2-e^2 + \frac{3r^4_H}{R^2}}{2r^3_{H}\sqrt{f(r,m,e,R)}}.
\ee
As a result, the detector of the above Rindler-like motion would measure
the correct Hawking temperature through the relation of
$T=a_7/2\pi$ as follows
\be
\label{temp8} T=\frac{r^2_H-e^2+ \frac{3r_H^4}{R^2}}{4\pi
r^3_H\sqrt{f(r,m,e,R)}}, \ee 
in the GEMS approach. Then, the desired BH temperature is given by
\be
T_0=\sqrt{g_{00}}T=
\frac{r^2_H-e^2+ \frac{3r_H^4}{R^2}}{4 \pi r^3_H}.
\ee


It is by now well--known that entropy, which is the extensive companion 
of the temperature, is given by one quarter of the horizon area \cite{g2}. 
On the other hand, R. Laflamme \cite{lafl} showed that entropy seen by 
an accelerated observer in Minkowski space can be obtained from 
the consideration of the transverse area of a null surface on the wedge. 
This transverse area would 
diverge for otherwise unrestricted Rindler motion due to the integration
over the whole transverse dimensions. 
In an embedded higher dimensional flat space, however, 
since there are ``embedding" constraints, the resulting integral may not
be divergent and make entropy finite. 

Our RN--AdS case, where there are three additional dimensions in the
transverse area, $\int dz^2 \ldots dz^6$, is correspondingly 
subject to four constraints as follows
\begin{eqnarray}
&&  \label{constr}
(z^1)^2-(z^0)^2 =0, \\
&&  \label{constr2}
z^2=f_{1}(r),~ z^6=f_{2}(r),  \\
&& (z^3)^2 + (z^4)^2 + (z^5)^2 = r^2,
\end{eqnarray}
where $f_i(r)$ are explicitly given in Eq. (\ref{rna}).
Note that Eq. (\ref{constr}) leads to $r=r_H$.
Since the $z^2$ and $z^6$ integrals subject to the constraints
(\ref{constr2}), 
$\int dz^2 dz^6 \delta (z^2-f_1(r)) \delta(z^6-f_2(r))$, is unity, 
the remaining integrals of $z_i (i=3,4,5)$ well
reproduce the desired area $4\pi r_H^2$ of the $r=r_H$ sphere.
This ends the global flat embedding of the RN--AdS space giving 
the correct thermodynamics.

\section{Various Limiting Geometries}

Now, we are ready to analyze the various limiting geometries
through the successive truncation procedure of the parameters, $e$, or $R$ (or, both) in the
original curved space.

\subsection{RN limit}

Let us first consider the RN limit \cite{rn, ksy}, which is the case of
$R\rightarrow \infty$ in the metric (\ref{metric}),
\be
\label{rn-ori}
ds_4^2= f_e (r,m,e) dt^2 - f^{-1}_e (r,m,e) dr^2
        -r^2 (d\theta^2+\sin^2 \theta d\phi^2),
\ee
where
\be
f_e (r,m,e) = f(r,m,e,R\rightarrow\infty) \equiv 1 -
\frac{2m}{r}+\frac{e^2}{r^2}. \ee The global flat embedding
coordinates can be obtained either in the GEMS approach \cite{des}
starting from the RN metric, Eq. (\ref{rn-ori}), or in the limit
of $R\rightarrow \infty$ from Eq. (\ref{rna}) directly as
\begin{eqnarray}
\label{rn}
&& z^0 = k_H^{-1}\sqrt{f_e(r,m,e)}\sinh (k_Ht), \nonumber \\
&& z^1 = k_H^{-1}\sqrt{f_e(r,m,e)}\cosh (k_Ht), \nonumber \\
&& z^2 = \int dr \left
         (\frac{r^2(r_{+}+r_{-}) + r_{+}^2 (r+r_{+})}{r^2(r-r_{-})}
         \right)^{1/2}, \nonumber \\
&& z^3=r\sin\theta\cos\phi, \nonumber \\
&& z^4=r\sin\theta\sin\phi, \nonumber \\
&& z^5=r\cos\theta, \nonumber \\
&& z^6 = \int dr \left
         (\frac{4r^5_{+}r_{-}}{r^4(r_{+}-r_{-})^2}
         \right)^{1/2},
\end{eqnarray}
where the surface gravity is given by
$k_H=k_H(r_H,e,\infty)=(r_{+}-r_{-})/2r^2_{+}$ with the
outer horizon $r_{+} = r_H$, and $r_{\pm}=m\pm
\sqrt{m^{2}-e^{2}}$. 
In this limit, the $R$-dominant part of $z^{6}$ in
Eq. (\ref{rna}) vanishes and the resulting GEMS becomes exactly the
known $D=7$ RN one \cite{des}. Note that in the limit of
$R\rightarrow \infty$ the corresponding event horizon becomes the
usual RN one by rewriting the charge $e^2$ to $r_{+}r_{-}$.

Moreover, the relevant $D=7$ acceleration and the Hawking
temperature can be obtained either directly from Eqs. (\ref{acc8})
and (\ref{temp8}) by taking the limit of $R\rightarrow \infty$ and
replacing $e^2$ with $r_{+}r_{-}$, or from the Rindler-like
motion in the $D=7$ GEMS, Eq. (\ref{rn}), following a static trajectory ($r,
\theta, \phi = {\rm constrant}$) in the curved space as before,
\begin{eqnarray}
\label{acc7}
a_7&=& \{ (z^1)^2-(z^0)^2 \}^{-1/2}=
             \frac{r_{+}-r_{-}}{2r^2_{+}\sqrt{f_{e}(r,m,e)}}, \\
\label{temp7}
T&=&\frac{r^2_{+}-e^2}{4\pi r_{+}^{3}\sqrt{f_{e}(r,m,e)}}=
\frac{r_{+}-r_{-}}{4\pi r_{+}^{2}\sqrt{f_{e}(r,m,e)}}.
\end{eqnarray}

The entropy calculation of the RN is essentially the same as the previous
RN--AdS case. In this case there are three additional
dimensions, and four constraints, {\it i.e.,}
$(z^1)^2-(z^0)^2=0$ leads to $r=r_{+}$,
$z^2=f_1(r,R\rightarrow\infty)$, $z^6=f_2(r,R\rightarrow\infty)$
in Eqs. (\ref{rna}) and $(z^3)^2 + (z^4)^2 + (z^5)^2 = r^2$.
Thus, since the $z^2, z^6$ integrals,
$\int dz^2 dz^6 \delta(z^2-f_1(r)) \delta(z^6-f_2(r))$, are unity,
the remaining integrals give the desired area 4$\pi r_H^2$,
that of the corresponding $r=r_H$ sphere.

\subsection{Schwarzschild-AdS limit}

Secondly, the RN--AdS solution (\ref{rna}) is also easily
reduced to the Schwarzschild--AdS space,
which is the limiting case of $e\rightarrow 0$,
\be
\label{sads1}
ds_4^2 = f_R (r,m,R) dt^2 -f^{-1}_R (r,m,R) dr^2
       -r^2(d{\theta}^2+\sin^2\theta d{\phi}^2),
\ee
where
\be
f_{R}(r,m,R) = f(r,m,e=0,R) \equiv 1-\frac{2m}{r} + \frac{r^{2}}{R^{2}},
\ee
giving another $D=7$ GEMS with the vanishing $e$-sensitive part of $z^6$
in Eq. (\ref{rna}),
\begin{eqnarray}
\label{sads}
&& z^0 = k_H^{-1}\sqrt{f_{R}(r,m,R)}\sinh (k_Ht), \nonumber \\
&& z^1 = k_H^{-1}\sqrt{f_{R}(r,m,R)}\cosh (k_Ht), \nonumber \\
&& z^2 = \int dr \frac{R^3+Rr^2_H}{R^2+3r^2_H}
          \sqrt{\frac{r_H (r^2 + rr_{H} + r_{H}^2)}
          {r^3(r^2 + rr_{H} + r_{H}^2+R^2)}}, \nonumber \\
&& z^3=r\sin\theta\cos\phi, \nonumber \\
&& z^4=r\sin\theta\sin\phi, \nonumber \\
&& z^5=r\cos\theta, \nonumber \\
&& z^6 = \int dr \sqrt{\frac{(R^4+10R^2r_H^2+9r^4_H)
                        (r^2 + rr_{H} + r_{H}^2)}
                           {(R^2+3r_H^2)^2 (r^2 + rr_{H} + r_{H}^2+R^2)}}.
\end{eqnarray}
The surface gravity, $k_H=k_H(r_H,0,R)=(R^2+3r_H^2)/2r_H R^2$, is now
either obtained at the root $r_H$ of $f_R(r,m,R)\mid_{r=r_H}=0$,
or reduced directly from the Eq. (\ref{sgrna}) with $e=0$. This
seemingly complicated embedding space is firstly obtained in Ref.
\cite{des}, and we have also reached to the exactly same results
by the systematic reduction process from Eq. (\ref{rna}).

On the other hand, similar to the RN limit case, we directly 
obtain the Hawking temperature from Eqs. (\ref{acc8}) and (\ref{temp8})
by taking the limit of $e\rightarrow 0$ as follows
\be
T=\frac{a_7}{2\pi}= \frac{1+ \frac{3r^2_H}{R^2}}{4\pi r_H\sqrt{f_{R}(r,m,R)}},
\ee
which again equals to that calculated in \cite{brown}.

\subsection{Schwarzschild limit}

Thirdly, we can obtain the Schwarzschild limit without the cosmological
constant from the RN embedding of (\ref{rn}) with $e\rightarrow 0$ limit
or the Schwarzschild--AdS embedding of (\ref{sads}) with
$R\rightarrow \infty$ one.
As a result, it is successfully reduced to the $D=6$ flat GEMS as follows
\cite{fro},
\begin{eqnarray}
\label{schw}
&& z^0 = k^{-1}_H \sqrt{1-2m/r}\sinh (k^{-1}_Ht), \nonumber \\
&& z^1 = k^{-1}_H \sqrt{1-2m/r}\cosh (k^{-1}_Ht), \nonumber \\
&& z^2 = \int dr \sqrt{ r_H (r^2 + rr_H + r_H^2)/r^3}, \nonumber \\
&& z^3 = r\sin \theta\sin\phi, \nonumber\\
&& z^4=r\sin\theta\cos\phi, \nonumber \\
&& z^5=r\cos\theta,
\end{eqnarray}
where the event horizon is $r_H=2m$, and the surface gravity is
$k_H(r_H, 0,\infty)=1/2r_H$. 
Note that the analyticity of $z^2(r)$ in $r>0$ covers the region of $r<r_H$. Thus,
it should be cautioned that the use of incomplete embedding spaces, 
that cover only $r>r_H$ (as, for example, in \cite{ros}), 
will lead to observers there for whom there is no event horizon, 
no loss of information, and no temperature.

We then obtain the Hawking
temperature from Eqs. (\ref{acc7}) and (\ref{temp7}) by taking the
limit $e\rightarrow 0$ as follows
\begin{eqnarray}
&&T = \frac{a_{6}}{2\pi}=\frac{1}{8\pi m\sqrt{1-2m/r}},
                   \nonumber \\
&&T_0= \sqrt{g_{00}}T=\frac{1}{8\pi m}.
\end{eqnarray}

It seems appropriate to comment on a global flat embedding of
$D=4$ covering of the AdS,
\begin{equation}
\label{covering}
ds_4^2=(1+\frac{r^2}{R^2})dt^2-(1+\frac{r^2}{R^2})^{-1}dr^2
      -r^2(d\theta^2+\sin^2\theta d\phi^2),
\end{equation}
which corresponds to the case of $m\rightarrow 0$ in Eq.
(\ref{sads1}). In this case we cannot directly obtain a global
embedding from Eq. (\ref{sads}) since in the limit of
$m\rightarrow 0$ the surface gravity 
$k_H=1/2r_H=1/4m$ yields a
divergence. As discussed in Ref. \cite{des} in details, this
problem originally comes from the fact that there is no intrinsic
horizon of this space-time. However, there is of course the other
direct route to embed this space-time into the $D=5$ flat space-time
starting from the metric (\ref{sads1}) with $m=0$. Based on the
accelerating coordinate system, the correct temperature of $2\pi
T=(a^2-R^{-2})^{1/2}$ has been already found (For further details,
see Ref. \cite{des}).

Similar to the pure AdS case, we can directly analyze
the purely charged case with $m=0$ in the metric (\ref{rn-ori}) as
\begin{equation}
\label{charge}
ds_4^2=(1+\frac{e^2}{r^2})dt^2-(1+\frac{e^2}{r^2})^{-1}dr^2
      -r^2(d\theta^2+\sin^2\theta d\phi^2).
\end{equation}
As like the above $D=4$ covering of the AdS, 
this has also no event horizon. However, we can also embed this 
space-time into $D=5$ flat one 
in view of the accelerating coordinate frame as follows
\begin{eqnarray}
z^0&=&\sqrt{\rho^2-e^2}\sinh(\eta/e), \nonumber\\
z^1&=&\sqrt{\rho^2-e^2}\cosh(\eta/e), \nonumber\\
z^2&=&\rho\sinh\Phi\cos\theta, \nonumber\\
z^3&=&\rho\sinh\Phi\sin\theta, \nonumber\\
z^4&=&\rho\cosh\Phi,
\end{eqnarray}
where $-\infty < \eta,\Phi < \infty, -\pi < \theta < \pi$. 
While this coordinate patch only covers the region $\rho > e$, it can be
extended to the entire space similar to the four dimensional AdS
case \cite{des}. Then, we can easily obtain the temperature as
$2\pi T=(a^2-e^{-2})^{1/2}$ where the four acceleration $a$ is
given by $a=\rho^2/e^2(\rho^2-e^2)$.

Furthermore, if we take the limit $R\rightarrow \infty$ in the metric
(\ref{covering}), or $e\rightarrow 0$ in the metric
(\ref{charge}), we finally reach to the flat four
dimensional Minkowski space.

We have summarized all these results in Fig. 1 as a compact diagram,
which can be obtained through the systematic truncation procedure
from the seven dimensional flat embedding space.

\section{Summary}

In summary, we have shown that the Hawking thermal properties
map into their Unruh equivalents in the (5+2)-dimensional GEMS,
which is the lowest possible global embedding dimensions of the curved RN--AdS space.
The relevant curved space detectors become Rindler ones, whose temperatures and
entropies reproduce the originals. Our results of the RN--AdS in
the GEMS approach include the various limiting geometries, which
are the Reissner--Nordstr\"om, Schwarzschild--AdS, and Schwarzschild
space-times through the successive reduction procedure of the
parameters in the original space. 
As a result, the (5+2)-dimensional GEMS in Eq. (\ref{rna}) serves 
a unifying description of the global flat embedding of the various geometries.
It would be interesting to
consider other interesting applications of the GEMS, for example,
the rotating Kerr type geometries \cite{btz,kps,ker,hkp}.

\vskip 0.5cm

We are grateful for interesting discussions to Prof. G. W. Gibbons.
Y.W.K. acknowledges financial support from KOSEF, Y.J.P. from
the Ministry of Education, BK21 Project No. D-0055/99, and
K.S. for S.N.U. CTP and Ministry of Education for BK-21 Project.


\begin{figure}
\vskip .1cm
\hskip 2.0cm
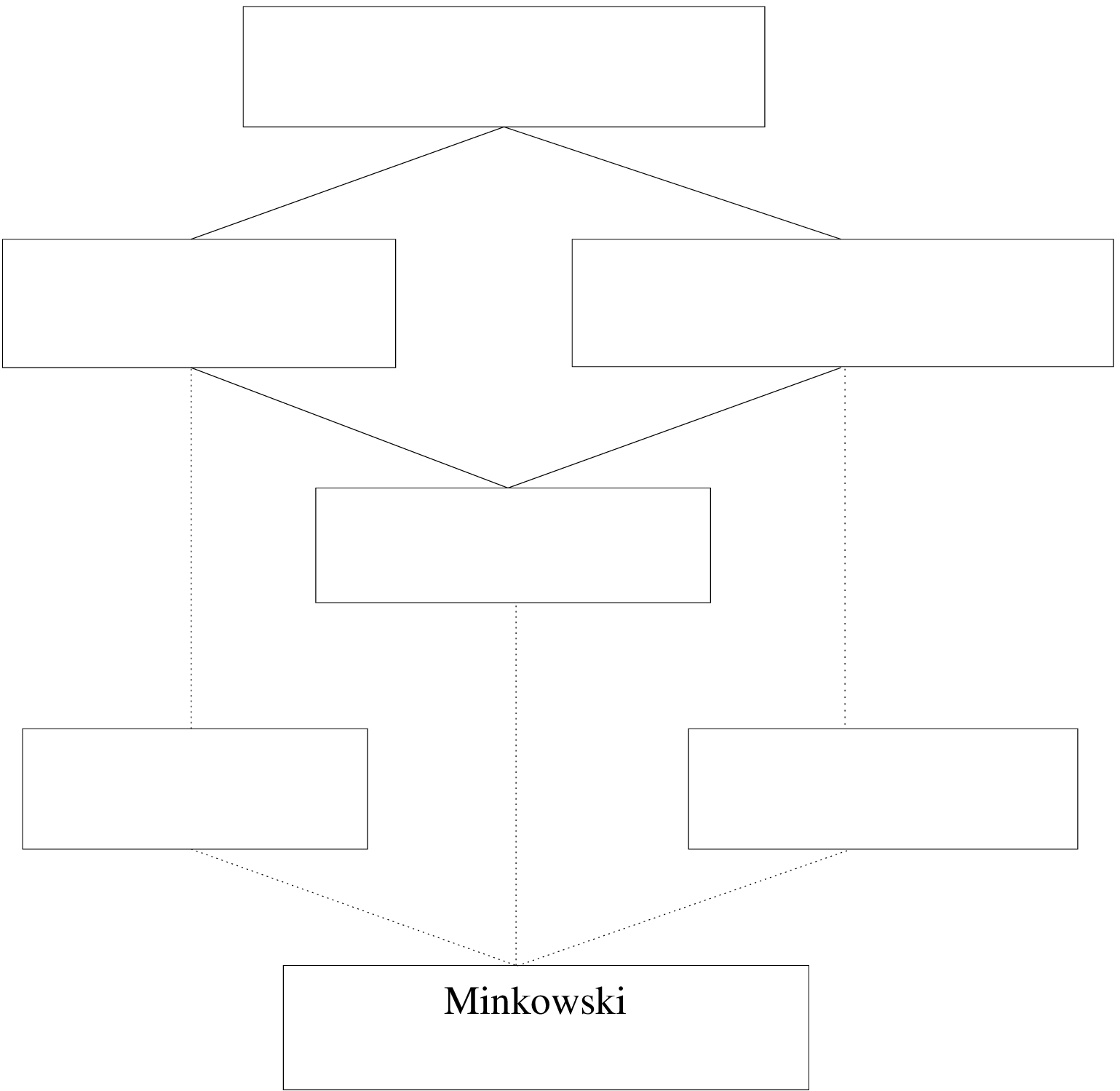
\caption{Various Limits of RN--AdS Embedding:
Trunction connected by solid lines means that through the direct
parameter reductions it is possible to obtain all thermodynamic 
quantities in the lower dimensional embedding system, while 
truncation connected by dotted lines means that these parameter reductions 
are only possible at the level of metric.
\hfill
\label{fig:ads.eps}}
\end{figure}

\end{document}